\documentclass[a4paper,11pt]{article}
\usepackage{pos}
\usepackage{float}

\usepackage[english]{babel}

\title{$R$-dependence of jet observables with JEWEL+v-USPhydro}

\author*[a]{Leonardo Barreto}
\author[a]{Fabio M. Canedo}
\author[a]{Maria M. M. Paulino}
\author[b]{Jacquelyn Noronha-Hostler}
\author[b]{Jorge Noronha}
\author[a]{Marcelo G. Munhoz}

\affiliation[a]{Instituto de F\'{i}sica, Universidade de S\~{a}o Paulo,\\
C.P. 66318, 05315-970 S\~{a}o Paulo, SP, Brazil}

\affiliation[b]{Illinois Center for Advanced Studies of the Universe Department of Physics, University of Illinois at Urbana-Champaign,\\
Urbana, IL 61801, USA}

\emailAdd{leonardo.barreto.campos@usp.br}
\emailAdd{fabio.canedo@usp.br}
\emailAdd{maria.paulino@usp.br}
\emailAdd{jnorhos@illinois.edu}
\emailAdd{jn0508@illinois.edu}
\emailAdd{munhoz@if.usp.br}

\graphicspath{ {./images/} }

\newcommand{\tr}{$\rm T_{R}ENTo$}

\newcommand{\qgp}{Quark-Gluon Plasma}

\newcommand{\tv}{\tr+v-USPhydro}
\newcommand{\vu}{v-USPhydro}

\newcommand{\gb}{Glauber+Bjorken}

\newcommand{\jv}{JEWEL+v-USPhydro}

\abstract{The $R$-dependence of jet observables provides a new tool in understanding the interplay between the jet energy-loss mechanism and medium response in heavy-ion collisions. This work applies the Monte Carlo events generator JEWEL and PYTHIA, coupled with $\rm T_{R}ENTo$ initial conditions and the state-of-the-art (2+1)D v-USPhydro, for the simulation of jet distributions and substructure observables for lead-lead collisions at LHC energy scales. We present the jet nuclear modification $R_{AA}$ and anisotropic flow coefficients $v_{n=2,3}$ varying the jet cone radius $R$, in the context of anti-$k_T$ jets, in addition to leading subjet fragmentation. The calculations indicate the impacts of the hydrodynamic evolution and weakly-coupled medium response, given by recoils, on the distributions. Results are compared to experimental data in a wide range of jet $p_T$ and collision centrality, and displayed along large jets ($R \ge 0.6$) predictions.}

\FullConference{HardProbes2023\\
 26-31 March 2023\\
 Aschaffenburg, Germany\\}

\begin{document}
\maketitle

\section{Introduction}

The exploration of the jet resolution parameter $R$, denominated cone radius regarding the anti-$k_T$ algorithm, has been shown to be a useful tool to better understand the jet quenching and medium response mechanisms in relativistic heavy-ion collisions \cite{Barreto:2022ulg}. Measurements of $R$-dependent jet distributions and substructure observables provide a phenomenological opportunity to constrain models that incorporate the interplay between jets and the \qgp{} (QGP).

This study investigates the impact of a realistic medium simulation in jet observables by coupling the state-of-the-art event-by-event viscous hydrodynamic (2+1)D \vu{} code \cite{Noronha-Hostler:2013gga, Noronha-Hostler:2014dqa} with the well-established JEWEL (Jet Evolution with Energy Loss) event generator \cite{Zapp:2012ak}. Partonic evolution in JEWEL original simplistic media cannot replicate $R$-varying jet nuclear modification factor $R_{AA}$ results for both low \cite{ALICE:2023waz} and high-$p_T$ \cite{CMS:2021vui} ranges, specially for large jets ($R \ge 0.6$). Thus $R$-dependent calculations provide a great scenario to assess the model improvement by the \vu{} addition.

\section{The JEWEL+v-USPhydro Model}

JEWEL 2.2.0 is a Monte-Carlo generator for parton showers based on the BDMPS-Z formalism with medium interaction and coherent gluon emission. It employs PYTHIA 6.4 \cite{Sjostrand:2006za} to generate the initial hard scattering and the hadronization process. 

\begin{figure}[b!]
    \centering
    \includegraphics[keepaspectratio, width=0.8\linewidth]{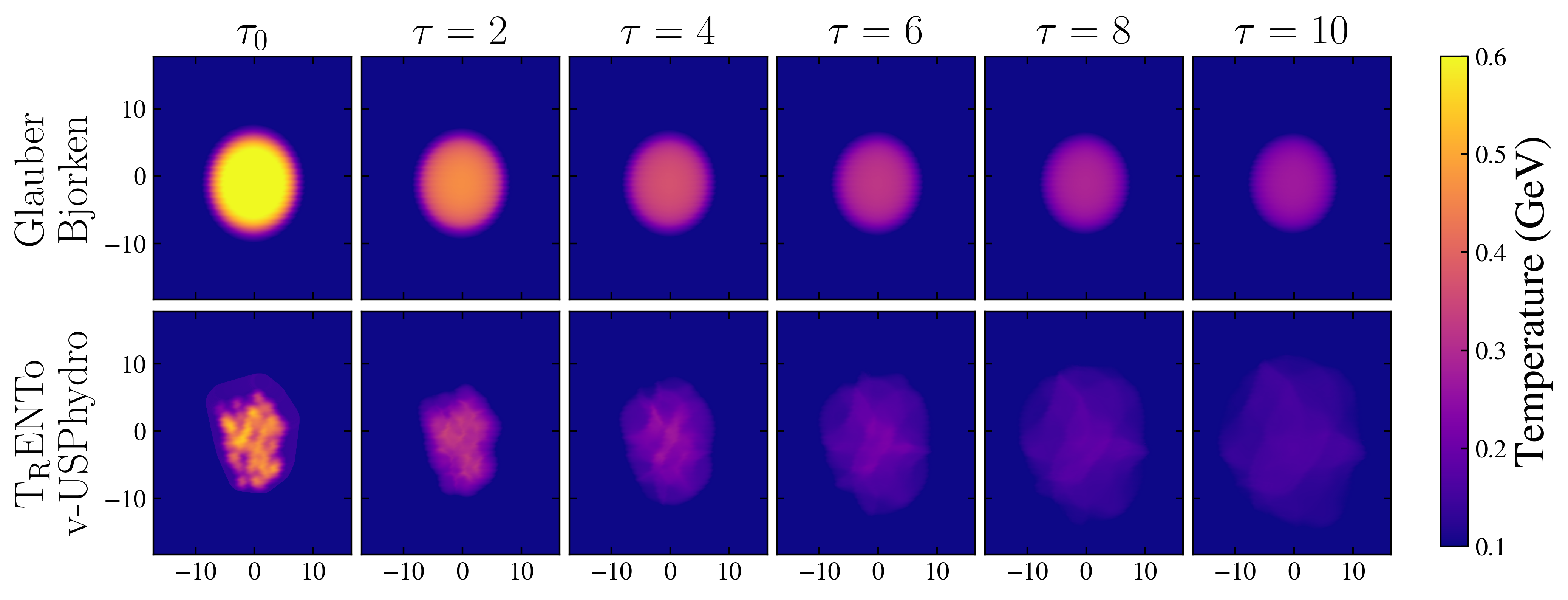}
    \caption{Temperature evolution using \gb{} (top row, JEWEL original medium) and \tv{} (bottom row) simulations for random central PbPb collisions at 5.02 TeV. Length and proper time scale in fm.}
    \label{fig:hydro_evolve}
 \end{figure}

For the medium description, JEWEL has a smooth and static Glauber approach with Bjorken-only expansion and no event-by-event fluctuations. The modification substitutes this simplistic picture with \tr{} initial conditions \cite{Moreland:2014oya} along the \vu{} model, which consider a transverse expansion with shear viscosity and local velocity. Figure \ref{fig:hydro_evolve} displays a comparison between both simulations. Further details of the implementation can be found in Refs. \cite{Canedo:2020xzf, Barreto:2022ulg}. 

We present the calculations with and without recoiling scattering centers (SC) with the 4MomSub procedure \cite{KunnawalkamElayavalli:2017hxo} to subtract the initial SC 4-momenta from the final distributions. We treat each option as independent models and tune them separately to ATLAS central $R_{AA}$ results to eliminate predictability biases \cite{Barreto:2022ulg, Barreto:2021fbt}. Results are provided with statistical uncertainties only.

We emphasize that JEWEL parton shower evolution and medium interaction formalism are not modified, only a new medium interface is provided with reinterpreted hydrodynamic information.

\section{Results}

Full calculations with experimental comparisons of the nuclear modification factor with fixed $R=0.4$ and elliptic and triangular anisotropic flow coefficients with $R=0.2$ for multiple centralities can be found in Ref. \cite{Barreto:2022ulg}. We shall focus exclusively on the $R$-dependence of these observables, with the inclusion of leading subjet fragmentation to better understand the $p_T$ distribution inside the jets.


The nuclear modification factor $R_{AA}$ is one of the main tools to quantify jet modification as the average jet yield is suppressed due to jet energy loss in AA systems. Given the differential jet yield $\frac{d^2N}{dp_T dy}$ normalized by the number of events $N_{\text{evt}}$ events, the observable is defined by
  \begin{equation}
    R_{AA}(p_T) = \frac{1}{\langle N_{\text{coll}} \rangle}\frac{\frac{1}{N_{\text{evt}}} \frac{d^2N}{dp_T dy}\big|_{AA}}{\frac{1}{N_{\text{evt}}} \frac{d^2 N}{dp_T dy}\big|_{pp}},
  \end{equation}
where the average number of binary nucleon collision $\langle N_{\text{coll}} \rangle$ is set to 1 for all events in the JEWEL framework.


Anisotropic flow coefficients for jets originate from differences in partonic evolution distances inside the \qgp. Initial condition eccentricities and path-length dependent energy-loss mechanism in AA systems imply in the modification of the jets' azimuthal distribution. The $R$-dependence of these observables is of particular interest since one of the major obstacles of measuring large jets is the presence of the highly-anisotropic background generated in heavy-ion collisions, which contribution to jets increases with $R^2$ \cite{ALICE:2023waz}. Therefore a better understanding of jet anisotropies may bring insight into background subtraction methodologies.

\begin{figure}[b!]
    \centering
    \includegraphics[keepaspectratio, width=0.8\linewidth]{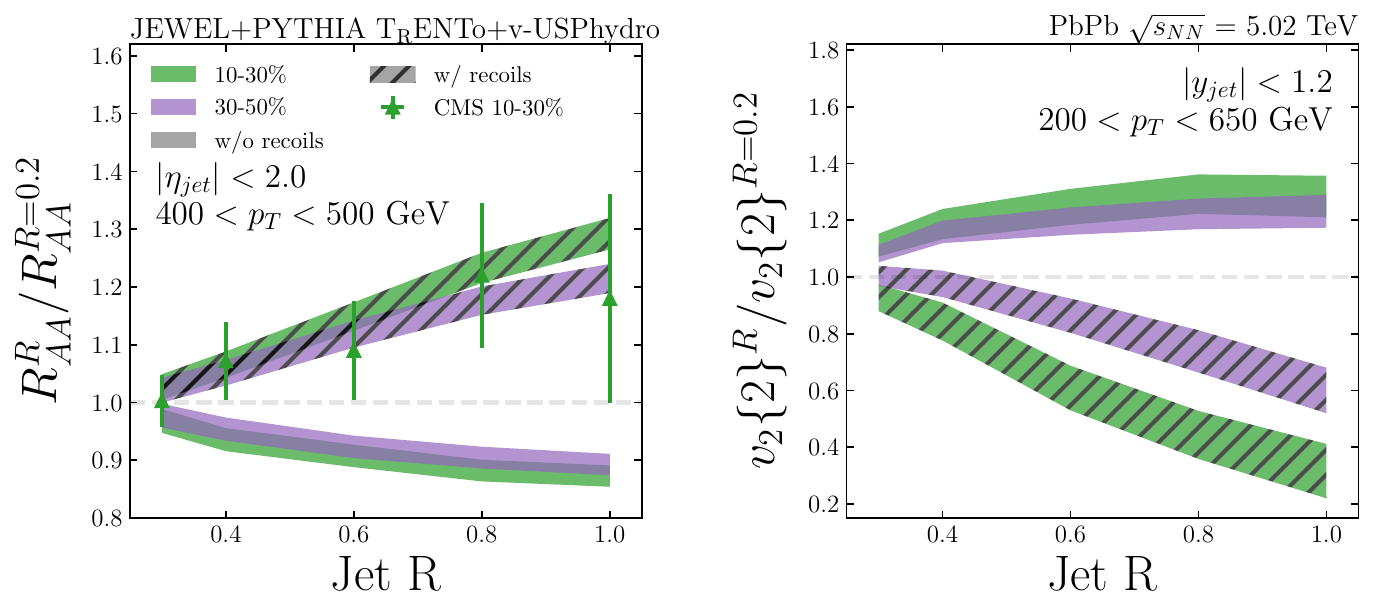}
    \caption{Integrated $R_{AA}$ (left), compared to CMS data \cite{CMS:2021vui}, and $v_2\{2\}$ (right) ratios dependence on the jet cone radius for 10-30\% and 30-50\% centralities. Kinematic cuts are presented in each panel.}
    \label{fig:ratio}
\end{figure}

Experimental results should be compared to the jet-soft correlation \cite{Betz:2016ayq}

\begin{equation}\label{eq:vncorr}
  v_n\{2\}(p_T) = \frac{\langle v_n^{\text{soft}}v_n^{\text{jet}}(p_T) \cos(n(\Psi_n^{\text{soft}} - \Psi_n^{\text{jet}}(p_T))) \rangle}{\sqrt{\Big \langle \left(v_n^{\text{soft}} \right)^2 \Big \rangle}},
\end{equation}
where definitions of the symmetry planes $\Psi_n$ and coefficients $v_n$ can be found in \cite{Barreto:2022ulg}.


In Figure \ref{fig:ratio}, we show the $R$-scaling ratio of $R_{AA}$ and $v_2\{2\}$, the curves indicate clear distinguished behavior due to medium response. The $R_{AA}$ is compared to 10-30\% CMS data \cite{CMS:2021vui} and indicates a significant improvement to JEWEL original model (results in the reference), being able to describe the experimental $R$ trend if recoils are considered. $v_2\{2\}$ has the exact opposite behavior from $R_{AA}$, with no further differences given the inclusion of recoils. 

\begin{figure}[t!]
    \centering
    \includegraphics[keepaspectratio, width=0.8\linewidth]{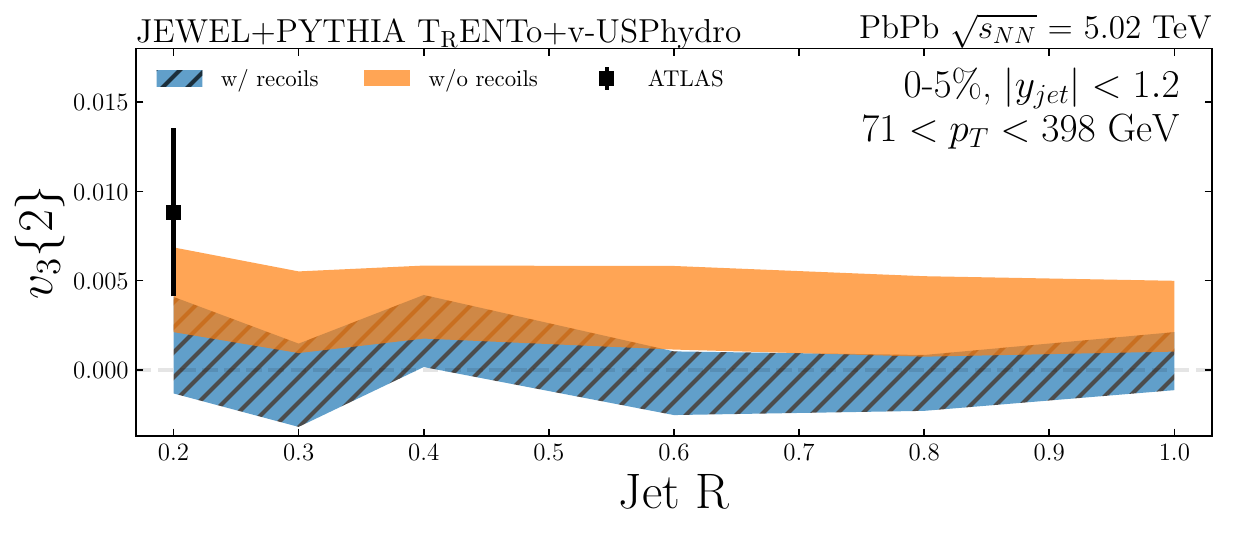}
    \caption{Integrated triangular flow $v_3\{2\}$ of jets dependence on jet cone radius compared to ATLAS results \cite{ATLAS:2021ktw} for 0-5\% centrality.}
    \label{fig:v3R}
\end{figure}

A closer look is taken into triangular flow in Figure \ref{fig:v3R} for central 0-5\% collisions. Both curves are lower than the ATLAS data \cite{ATLAS:2021ktw} that could be caused by a decorrelation effect, i.e. misalignment of $\Psi^{\text{soft}}$ and $\Psi^{\text{jet}}$, given a missing component in the jet-medium dialogue \cite{Barreto:2021fbt}. This is further indicated by $v_3\{2\}_{\text{recoils}} \sim 0$ as a recoiled SC does not interact with the medium after emission, thus increasing the decorrelation due to randomness in the $\Psi^{\text{jet}}$ distributions.

Lastly, we discuss the leading subjet fragmentation. Given a charged anti-$k_T$ with $R=0.4$, the jet clustering algorithm is reapplied to its constituents with a cone radius $r$ and the leading charged subjet is selected, defining its fraction of transverse momentum \cite{ALICE:2022vsz}
\begin{equation}
    z_r = \frac{p_T^{\text{ch subjet}}}{p_T^{\text{ch jet}}}.
\end{equation}

\begin{figure}
    \centering
    \includegraphics[keepaspectratio, width=0.8\linewidth]{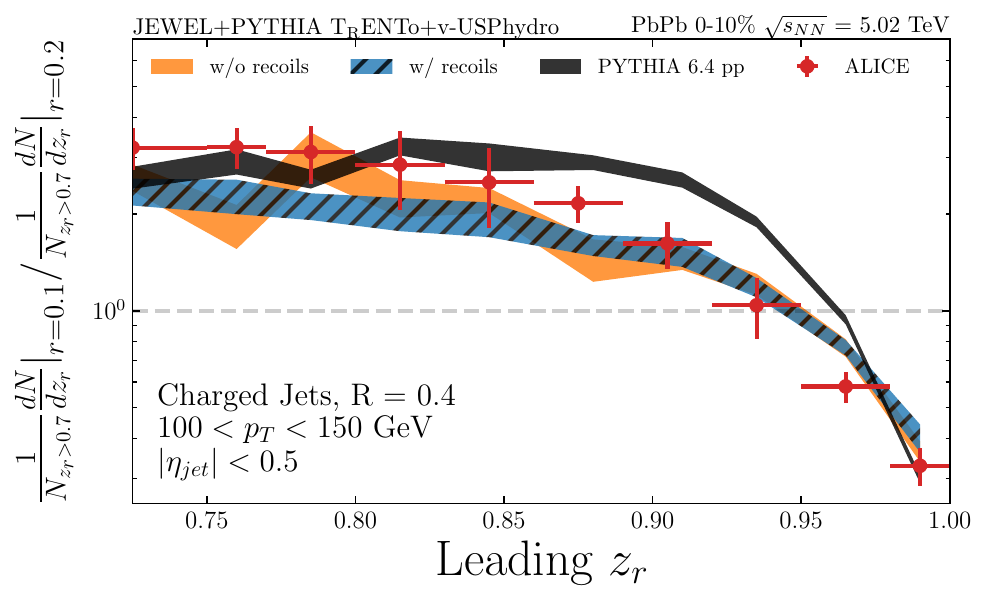}
    \caption{Ratio of leading subjet fragmentation distributions given the subjet radii $r=0.1$ and $0.2$ compared to ALICE PbPb 0-10\% results \cite{ALICE:2022vsz}. PYTHIA calculations for pp collisions (black) are displayed as a reference of jet behavior unmodified by \qgp{} interactions.}
    \label{fig:subjet}
\end{figure}

Figure \ref{fig:subjet} shows the ratio between $z_r$ distributions with $r=0.1$ and $0.2$ compared to ALICE central results \cite{ALICE:2022vsz}. Differently from JEWEL original calculations (in reference), the model describes the data within experimental uncertainties and little modification caused by the addition of recoils.

\section{Conclusions}

This work demonstrates a successful implementation of a realistic (2+1)D event-by-event hydrodynamic medium description in JEWEL. We observe an overall improvement in experimental data description when using \vu{} medium profiles and present predictions for large $R$ jet observables. 

The calculations show that inclusion of recoils seems to not be imperative for $R \le 0.4$ results, but is needed for large jets, in accordance with the expectation of medium response effects. The nuclear modification factor and elliptic flow coefficient display opposite behaviors in $R$-dependence, regardless the inclusion of recoils, while $v_3\{2\}$ is unchanged. There are indications of missing effects in the interplay of jets and medium, which shall be explored in future studies of the \jv{} model.

The \jv{} code is publicly available at \href{https://github.com/leo-barreto/USP-JEWEL}{github.com/leo-barreto/USP-JEWEL}, and the developed Rivet analyses at \href{https://github.com/leo-barreto/USPJWL-rivetanalyses}{github.com/leo-barreto/USPJWL-rivetanalyses}.

\section*{Acknowledgements}
L.B., F.M.C., M.M.M.P. and M.G.M. were supported by grant \#2012/04583-8, São Paulo Research Foundation (FAPESP). M.G.M. and F.M.C. acknowledge the support from Conselho Nacional de Desenvolvimento Científico e Tecnológico (CNPq) as well. J.N. is partially supported by the U.S.~Department of Energy, Office of Science, Office for Nuclear Physics under Award No. DE-SC0021301. J.N.H. acknowledges the support from the US-DOE Nuclear Science Grant No. DE-SC0020633, DE-SC0023861, within the framework of the Saturated Glue (SURGE) Topical Theory Collaboration, and the support from the Illinois Campus Cluster, a computing resource that is operated by the Illinois Campus Cluster Program (ICCP) in conjunction with the National Center for Supercomputing Applications (NCSA), and which is supported by funds from the University of Illinois at Urbana-Champaign. This study was financed in part by the Coordenação de Aperfeiçoamento de Pessoal de Nível Superior – Brasil (CAPES) – Finance Code 001.

\bibliographystyle{JHEP}
\bibliography{skeleton}

\end{document}